\documentclass[a4paper,11pt]{article}
\usepackage{jheppub}
\pdfoutput=1
\usepackage{hyperref,xcolor}
\usepackage{amsmath,amssymb}
\usepackage{graphicx}
\usepackage{xcolor}
\usepackage{mathtools}
\usepackage[utf8]{inputenc}
\usepackage[english]{babel}
\usepackage{bm}

\usepackage{layouts}
\usepackage{pgf,tikz}
\usetikzlibrary{arrows}
\usetikzlibrary[patterns]
\definecolor{zzttqq}{rgb}{0.6,0.2,0}
\definecolor{cqcqcq}{rgb}{0.75,0.75,0.75}
\definecolor{fftttt}{rgb}{1,0.2,0.2}
\definecolor{ffqqtt}{rgb}{1,0,0.2}
\definecolor{ffqqqq}{rgb}{1,0,0}

\newcommand{\Z}{\ensuremath{\mathbb{Z}}}

\newcommand{\rd}{\ensuremath{\mathrm{d}}}

\providecommand{\abs}[1]{\lvert#1\rvert}

\newcommand{\expv}[1]{\left\langle#1\right\rangle}

\newcommand{\be}{\begin{equation}}
\newcommand{\ee}{\end{equation}}
\newcommand{\benn}{\nonumber\begin{equation}}
\newcommand{\eenn}{\nonumber\end{equation}}
\def\bea{\begin{eqnarray}} \def\eea{\end{eqnarray}}
\def\beann{\begin{eqnarray*}} \def\eeann{\end{eqnarray*}}
\def\lsim{\raise0.3ex\hbox{$<$\kern-0.75em\raise-1.1ex\hbox{$\sim$}}}
\def\gsim{\raise0.3ex\hbox{$>$\kern-0.75em\raise-1.1ex\hbox{$\sim$}}}

\DeclareMathOperator\Tr{Tr}
\renewcommand{\Re}{{\rm Re}}
\renewcommand{\Im}{{\rm Im}}

\newcommand*\xbar[1]{%
  \hbox{%
    \vbox{%
      \hrule height 0.5pt 
      \kern0.3ex
      \hbox{%
        \kern-0.1em
        \ensuremath{#1}%
        \kern-0.1em
      }%
    }%
  }%
}

\graphicspath {{figures/}}

\title{Oscillating propagators in heavy-dense QCD}
\date{\today}
\author[a,1]{Oscar Akerlund,\note{Corresponding author.}}
\author[a,b]{Philippe de Forcrand}
\author[a]{Tobias Rindlisbacher}
\affiliation[a]{Institut f\"ur Theoretische Physik, ETH Z\"urich, CH-8093 Z\"urich, Switzerland}
\affiliation[b]{CERN, Physics Department, TH Unit, CH-1211 Geneva 23, Switzerland}

\emailAdd{oscara@itp.phys.ethz.ch}

\abstract{
Using Monte Carlo simulations and extended mean field theory calculations we show that the $3$-dimensional $\Z_3$ spin model
with complex external fields has non-monotonic spatial correlators in some regions of its parameter space. This model serves
as a proxy for heavy-dense QCD in $(3+1)$ dimensions. Non-monotonic spatial correlators are intrinsically related to a complex
mass spectrum and a liquid-like (or crystalline) behavior. A liquid phase could have implications for
heavy-ion experiments, where it could leave detectable signals in the spatial correlations of baryons.
}

\begin{document}
\maketitle
\flushbottom

\section{Introduction}\noindent
Oscillating spatial propagators have been the subject of several studies and carry important information about the underlying physics.
Patel~\cite{Patel:1983sc,Patel:1983qc} has argued that many-body correlations among the hadrons produced in heavy-ion collisions
may be oscillatory and has shown how those signals can be related to hadronization properties of the quark-gluon plasma (QGP)~\cite{Patel:2011dp}.
The underlying idea is that the QGP can be described as a network of quarks and of flux tubes into which the gluonic degrees of
freedom are concentrated. The flux tubes are assumed to interact mainly via three-point vertices, from here on called junctions,
where three flux tubes join together to form an $SU(3)$ singlet. It has been suggested that this system behaves like a liquid with
spatially oscillating two-body correlation between junctions and this structure might remain as the QGP hadronizes. This would be the case
if the string network breaks up via pair production rather than via coalescence of junctions. If that happens, then
the oscillatory signature should persist also in the two-body correlations of transversely outgoing hadrons.

Another situation with oscillating spatial correlation functions is in a possible crystalline phase in the QCD phase diagram,
which may occur at high density and low temperature. The existence of such a phase is supported by the exact solution of
the $(1+1)$-dimensional Gross-Neveu model at high density~\cite{Thies:2006ti,deForcrand:2006zz}. While the system described above may show
liquid-like correlations, i.e. exponential decay modulated by a cosine, the signature of a crystalline phase would be a purely
trigonometric correlation function.

When we talk about liquid-like behavior above and below, we have in mind a system where the spatial correlation functions are exponentially damped,
but with an oscillating modulation. The typical example of such behavior is the hard spheres model~\cite{Hoover:1968}: below the jamming transition
characteristic of the solid phase, one observes a liquid phase where spheres like to form spherical shells, causing oscillations in the
density-density correlation. However, both liquids and gases are fluids and
they are typically analytically connected through a cross-over, like in the case of water. A rigorous distinction is therefore ambiguous, although
in the presence of a first order transition, it is of course easy to identify the liquid as the denser, less compressible state.

To understand better when to expect such non-monotonic behavior, Ogilvie et al. have in a series of
papers~\cite{Meisinger:2010be,Nishimura:2014kla,Nishimura:2015lit} studied models which break charge conjugation $\mathcal{C}$,
but remain invariant under the combined action of $\mathcal{C}$ and complex conjugation $\mathcal{K}$. QCD at
nonzero chemical potential $\mu$ has this property, but also simpler models like the Polyakov-Nambu-Jona Lasinio (PNJL)
model with nonzero $\mu$, $SU(3)$ (Polyakov loop) spin models with nonzero $\mu$, and even the three-state Potts model
with nonzero $\mu$ have the same property. So before tackling full QCD one can hope to learn the implications of this
symmetry pattern from simpler models, which may even in some cases be mapped to limiting cases of QCD itself. It is well
known that for QCD, the expectation value of the Polyakov loop differs from the expectation value of its Hermitian conjugate,
$\expv{\Tr_FL}\neq\expv{\Tr_FL^\dagger}$ when the chemical potential $\mu$ is nonzero. However, the free energies are real
because of the $\mathcal{C}\mathcal{K}$ symmetry. As a further consequence of the breaking of $\mathcal{C}$, the transfer matrix
$T$ is not Hermitian, which means that the eigenvalues are not all necessarily real. Because of the invariance under
$\mathcal{C}\mathcal{K}$, however, if $\lambda$ is an eigenvalue of $T$, then so is $\lambda^*$, i.e. the eigenvalues are
either real or occur in complex conjugate pairs. This is interesting because it implies, in the case where complex eigenvalues
occur, that the Polyakov loop correlator is non-monotonic.

The dependence of a correlation function on the eigenvalues of the transfer matrix is most easily demonstrated with a one-dimensional
lattice model, but generalizes to any correlator with fixed momenta in the orthogonal directions. Let $\phi_i,\; i=1,\ldots,N,$ be a field living on
a circle with $N$ sites. If $T$ is the transfer matrix connecting neighboring sites, then the correlation function of $\phi$ is given by
\begin{equation}
  \label{eq:corr_1d}
  \expv{\phi(x)\phi(0)} = \frac{\Tr\left(T^{N-x}\phi{}T^x\phi\right)}{\Tr\left(T^N\right)} =
  \frac{\Tr\Big(\Lambda^{N-x}\overbrace{P^{-1}\phi{}P}^{\tilde\phi}\Lambda^x\overbrace{P^{-1}\phi{}P}^{\tilde\phi}\Big)}{\Tr\left(\Lambda^N\right)},
\end{equation}
where $\left(P^{-1}TP\right)_{ij}=\Lambda_{ij} = \lambda_i\delta_{ij}$ is the diagonalized transfer matrix with the eigenvalues $\lambda$ of $T$
sorted in magnitude such that
$\Re{\lambda_0}\geq\Re{\lambda_1}\geq\cdots$ and $\tilde{\phi}_{ij}$ is $\phi$ in the eigenbasis of $T$. For simplicity we assume
a discrete spectrum in the description below. We also consider the
$N\to\infty$ limit. In general three scenarios are possible\footnote{The eigenvalues of the transfer matrix are either real or come in complex conjugate
pairs, since the model is invariant under the simultaneous action of charge and complex conjugation.}. Firstly, all eigenvalues can be real
and the correlator is a conventional, exponentially decaying function,
\begin{equation}
  \label{eq:corr_real}
  \expv{\phi(x)\phi(0)} = \sum_n\abs{\tilde{\phi}_{0n}}^2\left(\frac{\lambda_n}{\lambda_0}\right)^x =
  \abs{\tilde{\phi}_{00}}^2+\abs{\tilde{\phi}_{01}}^2e^{-m_1x} + \mathcal{O}(e^{-m_2x}).
\end{equation}
Here we parametrize $\lambda_n/\lambda_0=e^{-m_n}$, with $m_n\geq0$.
Secondly, if the largest eigenvalue is real and the next two are
a complex-conjugate pair\footnote{Strictly speaking, there can be more real eigenvalues above the complex-conjugate pair,
  with a consequently weaker oscillation in the correlator. We do not treat that case separately here.}, then the correlator also decays exponentially but is modulated by a cosine,
and the system behaves as a liquid,
\begin{equation}
  \label{eq:corr_liquid}
  \expv{\phi(x)\phi(0)} \approx \abs{\tilde{\phi}_{00}}^2 + \abs{\tilde{\phi}_{01}}^2\left(\left(\frac{\lambda_1}{\lambda_0}\right)^x+
    \left(\frac{\lambda_1^*}{\lambda_0}\right)^x\right) = \abs{\tilde{\phi}_{00}}^2 + 2\abs{\tilde{\phi}_{01}}^2e^{-m_Rx}\cos{}m_Ix,
\end{equation}
where $\lambda_1/\lambda_0=e^{-m_R+im_I}$.
Finally, if the eigenvalue with the largest real part is part of a conjugate pair $\lambda_1/\lambda_0=e^{im_I}$, then the correlator is a pure
trigonometric function and a crystalline behavior is observed,
\begin{align}
  \label{eq:corr_crystal}
  \expv{\phi(x)\phi(0)} &\approx \frac{\lambda_0^{N-x}\sum_n\abs{\tilde{\phi}_{0n}}^2\lambda_n^x+(\lambda_0^*)^{N-x}\sum_n\abs{\tilde{\phi}_{1n}}^2\lambda_n^x}
  {\lambda_0^N+(\lambda_0^*)^N}\nonumber\\
  &\approx \frac{e^{-im_IN/2}\left(\abs{\tilde{\phi}_{00}}^2+\abs{\tilde{\phi}_{01}}^2e^{im_Ix}\right)+
    e^{im_IN/2}\left(\abs{\tilde{\phi}_{10}}^2e^{-im_Ix}+\abs{\tilde{\phi}_{11}}^2\right)}{2\cos(m_IN/2)}\\
  &=\abs{\tilde{\phi}_{00}}^2+\frac{\cos\left(m_Ix-m_IN/2\right)}{\cos(m_IN/2)}\abs{\tilde{\phi}_{01}}^2,\nonumber
\end{align}
of the form $\abs{\tilde\phi_{00}}^2+A\cos(m_Ix+\theta)$, which reveals long-range
order for arbitrarily large system size $N$. If $m_I$ is small in units of the lattice spacing then one oscillation spans several
lattice spacings and has nothing to do with the underlying structure of the lattice. For a continuous spectrum, the above categorization is
still valid. In case one, the eigenvalues are all distributed on the real line whereas in case two, the eigenvalues branch off into the complex plane
somewhere below the largest eigenvalue. The transition between cases one and two occurs at a so called disorder line. Case three is obtained when
the branching point reaches the largest real eigenvalue.

All these three cases have been found by Ogilvie et al~\cite{Meisinger:2010be}
in $1$-dimensional models, where the complete phase diagram can be obtained using transfer-matrix methods. Such a $1$-dimensional model can
for example serve as a dimensionally reduced effective models of $1+1$-dimensional QCD at finite temperature. Recently~\cite{Nishimura:2015lit}
it has been proposed that also higher dimensional models show these characteristics, based on the fact that the $1$-dimensional solution
can be seen as the first order in a character expansion. It has, however, to our knowledge, not been demonstrated with first-principles
lattice simulations that this is actually the case.

As mentioned above, it has been suggested~\cite{Patel:2011dp} that the conditions in the fireball after a heavy-ion collision might be such that
the baryon-number correlations have an oscillatory character. This conjecture is based on an effective flux-tube model
introduced in~\cite{Patel:1983sc,Patel:1983qc} which can be mapped into an $XY$-model with external magnetic fields which break
charge symmetry, such that it falls in the same category of models discussed above. Another flux-tube model, which can be
mapped into a three-state Potts model, is treated in~\cite{Condella:1999bk}. In general, the Hamiltonian and partition function
for such a flux-tube model are given by
\begin{equation}
  \label{eq:ham_fluxtube}
  H = \sigma\sum_{x,\nu}\abs{l_{x,\nu}} + m\sum_x\abs{q_x} + v\sum_x\abs{j_x},\quad Z = \sum_{\mathclap{\{l_{x,\nu},q_x,j_x\}}}e^{-\beta(H-\mu \sum_x q_x)},
\end{equation}
where $l_{x,\nu}$ denote flux tubes with string tension $\sigma$ living on the links, $q_x$ denote quarks with mass $m$ and chemical
potential $\mu$ living on the sites and $j_x$ denote junctions with vertex energy $v$ living on the sites. All occupation
numbers are integer valued and, depending on their allowed range and on whether $v$ is zero or nonzero, the model can be mapped to
either an $XY$ model (for $v\neq0$) or a $Z_N$ spin model (for $v=0$). The junctions $j$ call for further explanation. In $SU(N)$,
they are related to the invariant
$\epsilon$-tensor, i.e. $N$ flux lines emanating from $N$ (anti-)quarks join at a junction and form an $SU(N)$ singlet, and thus the
(anti-)quarks together with the flux lines are identified with a (anti-)baryon.

In this report we study the $\mathbb{Z}_3$ spin model with nonzero chemical potential $\mu$ in 1 and 3 dimensions and show that
a complex mass spectrum can occur in both cases. The rest of the report is organized as follows. In section~\ref{sec:model} we
define the model and review the solution in 1 dimension using the transfer matrix as well as the approximate solution in any dimension
using Extended Mean Field Theory (EMFT)~\cite{Akerlund:2014mea}. In section~\ref{sec:results} we present our Monte Carlo results
and then we draw our conclusions in section~\ref{sec:conclusions}.

\section{Model}\noindent
\label{sec:model}
The model we will be studying is the three-states Potts model with nonzero chemical potential or, more accurately, the $\Z_3$ spin model with
complex external fields\footnote{It may be worth pointing out that this type of model is often called a 3-state Potts model.
  This is not entirely accurate since the $\mathbb{Z}_3$ spin model~\eqref{eq:S_potts} is only equivalent to a 3-state Potts model if $h_I=0$.}.
In $d$ dimensions, it can be seen as the crudest approximation of $(d+1)$-dimensional QCD in the static-dense limit.
The action is given by
\begin{equation}
  \label{eq:S_potts}
  S = -\beta\sum_{\expv{i,j}}\left(P_iP^\dagger_j + P^\dagger_iP_j\right) - 2\sum_i\left(h_R\Re{}P_i + ih_I\Im{}P_i\right),
\end{equation}
and the $\mathbb{Z}_3$ spins $P\in\{1,e^{i\frac{2\pi{}}{3}},e^{-i\frac{2\pi{}}{3}}\}$ at each site represent the center of the Polyakov loops $\Tr_FL$.
The usual interpretation of the external fields is $h_R = e^{-M/T}\cosh(\mu/T), h_I = e^{-M/T}\sinh(\mu/T)$, where $M$ and $\mu$ are
the mass and chemical potential of the quarks respectively~\cite{Alford:2001ug}, but as mentioned in the introduction, it is also possible to map
$(\beta,h_R,h_I)$ of eq.~\eqref{eq:S_potts} into $(\sigma,m,\mu)$ of eq.~\eqref{eq:ham_fluxtube}, as described in~\cite{Condella:1999bk} (see eqs.~(14-18)).
We will primarily use the first mapping but will evaluate the results also in the light of the second one. Note, however, that the mapping between
eq.~\eqref{eq:ham_fluxtube} and eq.~\eqref{eq:S_potts} is not possible for all parameter values (see Fig.~\ref{fig:pd_3d_z3_emft}).

In the formulation \eqref{eq:S_potts} the action is complex, and the model clearly suffers from a sign problem, but as long as $h_R,h_I\in\mathbb{R}$
and $h_R>\abs{h_I}$, which corresponds to the physical case of $M,\mu\in\mathbb{R}$, there exists a sign-problem-free
representation\footnote{This is essentially going back to the representation in terms of flux-tube variables}
that can be sampled by a worm algorithm. The model can however be interesting in its own right also in the unphysical region $h_I>h_R$,
but it is a shortcoming that it does not have a continuum limit in three dimensions, which could make it harder to clearly separate the
lattice spacing $a$ from the correlation length $\xi$ and in extension, the wavelength $\lambda$ of the oscillations we are looking for.
In one dimension the model can be solved for general external fields using a transfer-matrix method and we can use EMFT to obtain an approximate
solution in any number of dimensions.

\subsection{Transfer matrix}
\begin{figure}[htp]
\centering
\includegraphics[width=0.6\linewidth]{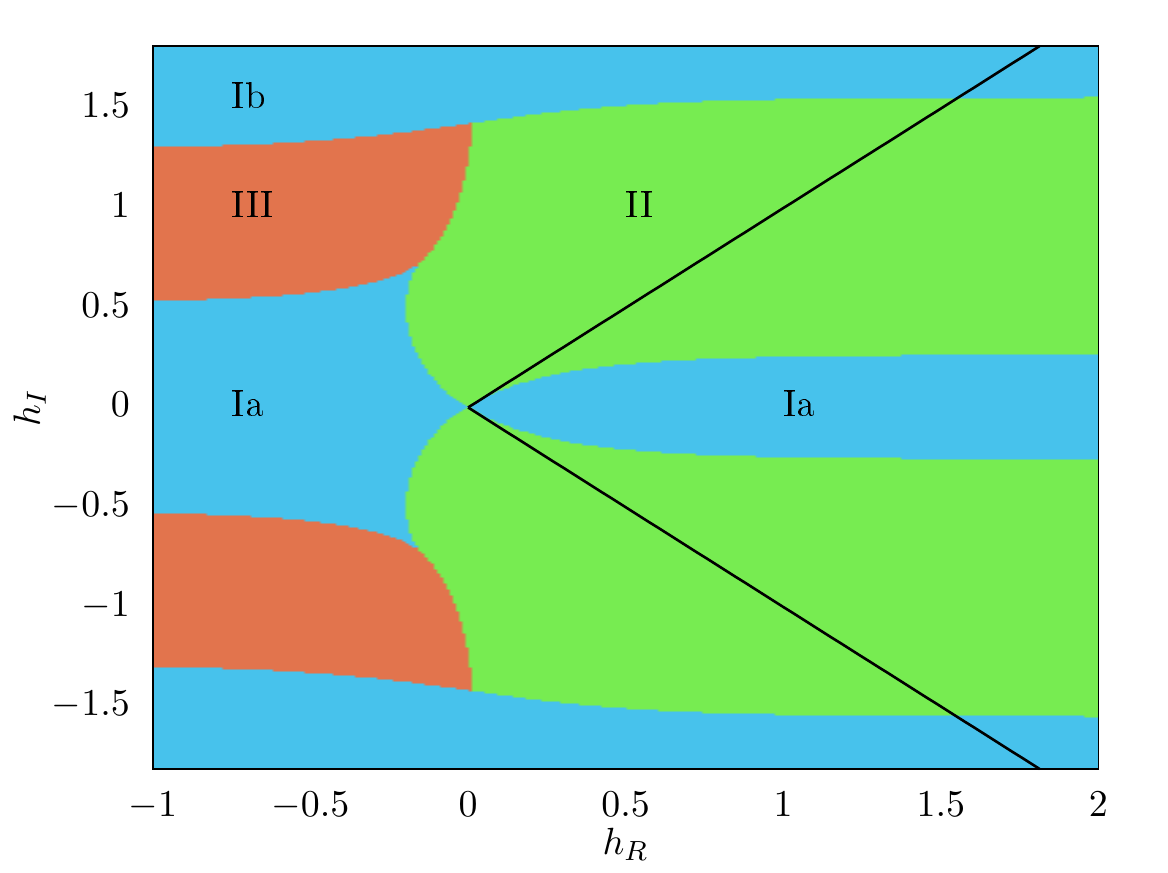}
\caption{Phase diagram of the $1d$ $\mathbb{Z}_3$ spin model in the $(h_R,h_I)$-plane for fixed $\beta=0.08$. The crystalline phase III is outside the
  region of parameter space where the worm algorithm can be applied but the liquid phase II is susceptible to lattice simulations. For a more
  detailed description of the phases see the text. The phase diagram is periodic in $h_I$ with period $\pi/\sqrt{3}$}
  \label{fig:pd_1d_z3_tm}
\end{figure}

\noindent
\label{sec:transfer_matrix}
In $1d$ the partition function of a chain of $N$ $\Z_3$ spins with periodic boundary conditions is given by
\begin{equation}
  \label{eq:Z_1d}
  Z = \Tr{}T^N, \quad T \propto \begin{pmatrix} e^{2\beta+2h_R} & e^{-\beta+\frac{h_R}{2}}e^{i\frac{\sqrt{3}h_I}{2}} & e^{-\beta+\frac{h_R}{2}}e^{-i\frac{\sqrt{3}h_I}{2}} \\
  e^{-\beta+\frac{h_R}{2}}e^{i\frac{\sqrt{3}h_I}{2}} & e^{2\beta-h_R}e^{i\sqrt{3}h_I} & e^{-\beta-h_R}\\
  e^{-\beta+\frac{h_R}{2}}e^{-i\frac{\sqrt{3}h_I}{2}} & e^{-\beta-h_R} & e^{2\beta-h_R}e^{-i\sqrt{3}h_I}\end{pmatrix} 
\end{equation}
where $T$ is the transfer matrix. It is easy to verify that the characteristic polynomial of $T$ is a cubic polynomial with real coefficients
so there are either three real roots or one real root and a pair of complex conjugate roots, as claimed above. For a given $\beta$, it is now
straightforward to determine the phase diagram which contains the three phases described in the introduction. The phase diagram at fixed
$\beta=0.08$ can be seen in Fig.~\ref{fig:pd_1d_z3_tm}. The color coding and labels are as follows: I (blue) marks the region where all eigenvalues
of $T$ are real. In Ia they are all positive and the connected correlator is a pure sum of exponentials. In Ib two eigenvalues are negative (the product of
all three, i.e. the determinant of $T$, is always positive) and the connected correlator is in general a sum of two oscillating functions with wavelength 2,
due to factors $(-1)^x$.
Depending on how the signs and magnitudes of the eigenvalues are distributed, this may or may not be detectable on a discrete lattice.
II (green) denotes the region where the largest eigenvalue is real and the other two are a complex conjugate pair. The connected correlator is a
cosine-modulated exponential, this is characteristic of a liquid. III (red) marks the region where the complex conjugate pair is larger in magnitude than the
real eigenvalue and the connected correlator at long distance is a pure trigonometric function, this is the long-range order characteristic of a crystal.
The two black lines bound the wedge
where $h_R>\abs{h_I}$ and mark the region where the flux-variables representation is sign-problem free and the worm algorithm can be used.
It is evident that the crystalline phase is out of reach of the worm algorithm but some parts of the liquid phase lie within the physical region
$h_R>\abs{h_I}$, so that the non-monotonic behavior of the connected correlator there can be reproduced by lattice simulations. Initially the transfer-matrix
method is only defined for integer separations but it is straight forward to extend it to any real separation via the matrix power-function.
In the liquid phase, the connected correlator is given by eq.~\eqref{eq:corr_liquid}, which is made periodic ($\exp\to\cosh$) at finite $N$ to obtain
\begin{align}
  \label{eq:corr_exact}
  \langle{}f(P(x))f(P(0))^\dagger\rangle_c = &\,a_f\left(\cosh\left(m_R\left(x-\frac{N}{2}\right)\right)\cos\left(m_I\left(x-\frac{N}{2}\right)\right)
    \cos\left(\phi_f\right)\right.\\
    &\phantom{a_f\Bigg(}\left.+\sinh\left(m_R\left(x-\frac{N}{2}\right)\right)\sin\left(m_I\left(x-\frac{N}{2}\right)\right)\sin\left(\phi_f\right)\right),\nonumber
\end{align}
where $f(P)$ is either $P,\Re{}P$ or $\Im{}P$. The parameters $a_f$ and $\phi_f$ can be calculated from the eigenvectors of $T$.
These functions can be directly compared to the correlators obtained by the worm algorithm and will serve as a consistency check for the algorithm
before going on to three dimensions where no exact results are available.

A comment about this ``phase diagram'' is in order. Actually, the different phases are not separated by phase transitions in the strict
sense; there is no singularity in the free energy anywhere in the $(h_R,h_I)$-plane, since the zeros of the characteristic polynomial of $T$
are smooth functions over the whole plane. Instead, the boundary of the different phases are \emph{disorder lines}, which mark a smooth change
in the characteristic of the correlator, for example from a non-oscillatory exponential decay to an oscillatory exponential decay. In general,
however, it is not necessarily so that the change from non-oscillatory to oscillatory behavior take place at a disorder line, it can also
occur at a first order transition, as is evident from for example the water-vapor transition.

\subsection{EMFT}\noindent
\label{sec:EMFT}
In more than one dimension, and especially in the physically interesting case of three dimensions, the transfer-matrix method is not practical
anymore. It is reasonable to assume that the structure of the phase diagram will remain~\cite{Nishimura:2015lit} but one is totally at
a loss when it comes to the exact location of the disorder lines. In the light of the one-dimensional results, it is unlikely that the
crystalline phase can be probed by lattice simulations, but one may hope to find evidence of a liquid phase. In this case the
three largest eigenvalues of the transfer matrix will be given by (up to a trivial overall multiplicative, real constant)
$\lambda_0 = 1, \lambda_1 = e^{-m_R-im_I}, \lambda_2 = e^{-m_R+im_I}$, where $m_R,m_I>0$ are real numbers chosen to paramterize the eigenvalues. 
The decay of the spin-spin correlator will thus be governed by $\expv{P(0)P^\dagger(r)}\sim e^{-m_Rr}\cos(m_Ir)$. It becomes clear that our
prospects for detecting this characteristic behavior of the correlator depend rather sensitively on $m_R$ and $m_I$; we require a
point in phase space where $m_R$ is not too large at the same time as $m_I$ is not too small, so that the first maximum in the correlator
occurs before the signal is too damped. Much time and effort can be saved by quickly,
albeit approximately, solving the model for extended regions of parameter space. Mean field theory is one candidate which falls short
since it does not give access to the mass spectrum. EMFT~\cite{Akerlund:2015fya} on the other hand does exactly that and is thus an apt choice.

\begin{figure}[htp]
\centering
\includegraphics[width=0.6\linewidth]{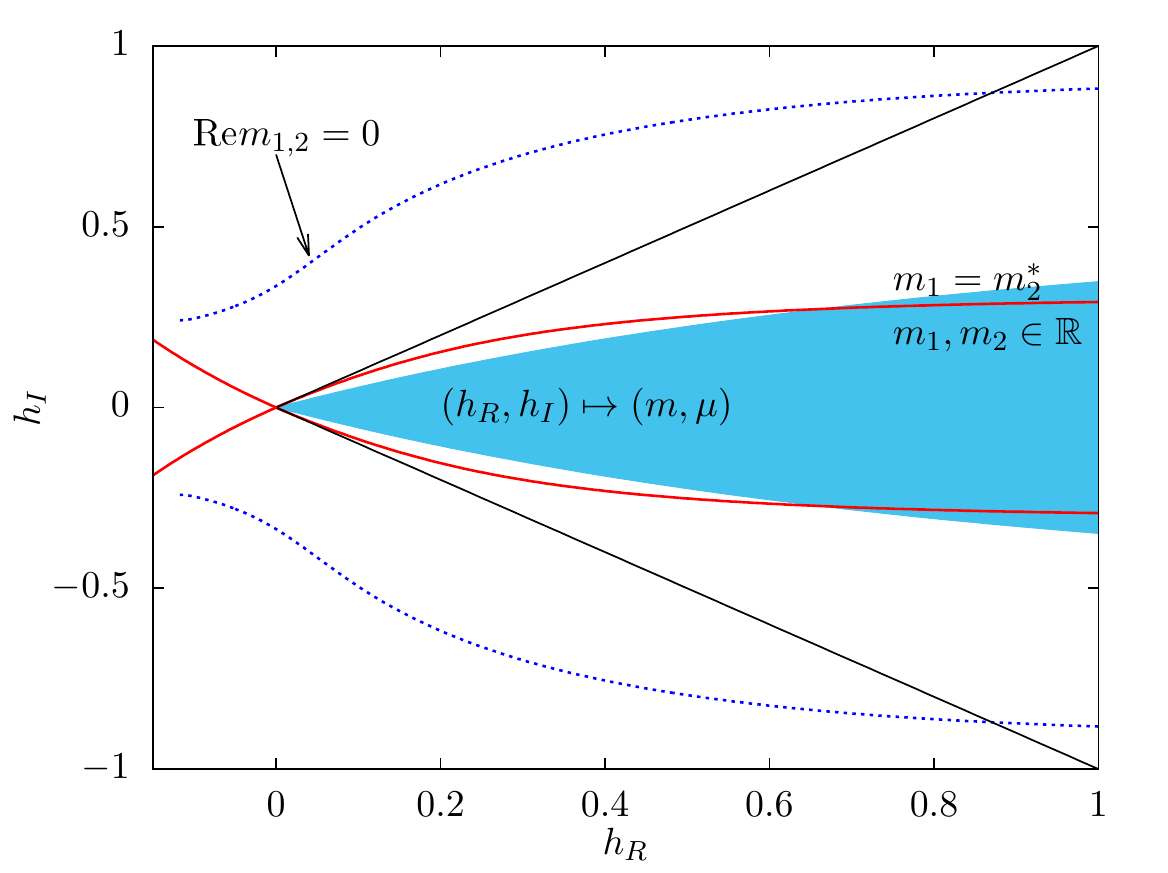}
\caption{Phase diagram of the $3d$ three-states Potts model at fixed $\beta=0.08$ obtained by EMFT. The thick red lines are disorder lines where
  the mass spectrum turns complex and on the dashed blue lines the real part of the mass vanishes. Those lines bound the region of convergence
  of EMFT. Inside the wedge bounded by the thin black lines the model~\eqref{eq:S_potts} is sign-problem free and the blue region marks the image
  of the map from the standard $\mathbb{Z}_3$ model to the flux-tube model of~\cite{Condella:1999bk}, eq.~\eqref{eq:ham_fluxtube} with $v=0$.}
  \label{fig:pd_3d_z3_emft}
\end{figure}

It will be useful to consider the real part $\Re{}P$ and the imaginary part $\Im{}P$ of the Potts spin $P$ as independent variables here.
Since the imaginary part of the action \eqref{eq:S_potts} is odd in $\Im{}P$, the expectation value of $i\Im{}P$ will be real and
we have $\expv{P}=\expv{\Re{}P}+\expv{i\Im{}P}\neq\expv{\Re{}P}-\expv{i\Im{}P}=\expv{P^\dagger}$. The $\mathbb{Z}_3$ spin $P$ is then decomposed
into its mean value and fluctuations around the mean,
\begin{align}
  \label{eq:P_split}
  P &= \expv{\Re{}P} + \delta\Re{}P+\expv{i\Im{}P} + i\delta\Im{}P,\\
  P^\dagger &= \expv{\Re{}P} + \delta\Re{}P-\expv{i\Im{}P} +-i\delta\Im{}P.\nonumber
\end{align}
We now formally integrate out all fields except the one at the origin and assume that this amounts to the introduction of effective couplings
for the bilinears $\delta\Re{}P\delta\Re{}P, \delta\Im{}P\delta\Im{}P$ and $\delta\Re{}P\delta\Im{}P$~\cite{Akerlund:2014mea}. The effective
EMFT action can then be written
\begin{align}
  \label{eq:S_EMFT}
  S_\text{EMFT} = &-\left(\Re{}P\right)^2\Delta_1-\left(\Im{}P\right)^2\Delta_2 - 2i\Re{}P\Im{}P\Delta_3 \nonumber\\
  &-2\Re{}P\left(h_R+\expv{\Re{}P}\left(2d\beta-\Delta_1\right)+\expv{i\Im{}P}\Delta_3\right)\\
  &-2i\Im{}P\left(h_I+\expv{i\Im{}P}\left(2d\beta-\Delta_2\right)-\expv{\Re{}P}\Delta_3\right).\nonumber
\end{align}
So far we have not assumed anything about the variables $P$, so the effective action above is generally valid for any action of the
form~\eqref{eq:S_potts}. For $P\in\mathbb{Z}_3$ the action can be simplified slightly by using $\left(\Im{}P\right)^2=1-\left(\Re{}P\right)^2$
and $\Re{}P=-\frac{1}{2}$ whenever $\Im{}P\neq0$. We then obtain
\begin{align}
  \label{eq:S_EMFT_simpl}
  S_\text{EMFT}&=-\left(\Re{}P\right)^2(\Delta_1-\Delta_2)-2\Re{}P\tilde{h}_R-\frac{2}{\sqrt{3}}i\Im{}P\tilde{h}_I,\\
  \tilde{h}_R &= h_R+\expv{\Re{}P}\left(2d\beta-\Delta_1\right)+\expv{i\Im{}P}\Delta_3,\\
  \frac{\tilde{h}_I}{\sqrt{3}} &= h_I+\expv{i\Im{}P}\left(2d\beta-\Delta_2\right)-\left(\expv{\Re{}P}+\frac{1}{2}\right)\Delta_3.
\end{align}
Defining $\log\gamma=-\frac{3}{4}\left(\Delta_1-\Delta_2\right)-3\tilde{h}_R$, it is straightforward to calculate all expectation values of the model
\begin{align}
  \expv{\Re{}P}&=\frac{1-\gamma\cos\tilde{h}_I}{1+2\gamma\cos\tilde{h}_I} &
  \expv{i\Im{}P}&=\frac{\sqrt{3}\gamma\sin\tilde{h}_I}{1+2\gamma\cos\tilde{h}_I} & & \\
  \expv{\left(\Re{}P\right)^2}&=\frac{1+\frac{1}{2}\gamma\cos\tilde{h}_I}{1+2\gamma\cos\tilde{h}_I} &
  \expv{\left(\Im{}P\right)^2}&=-\frac{\frac{3}{2}\gamma\cos\tilde{h}_I}{1+2\gamma\cos\tilde{h}_I} &
  \expv{i\Im{}P\Re{}P}&=\frac{-\frac{\sqrt{3}}{2}\gamma\sin\tilde{h}_I}{1+2\gamma\cos\tilde{h}_I}.\nonumber
\end{align}
It is obvious how to self-consistently determine the linear expectation values, whereas the bilinears may need some more explanation.
The details of their determination will reveal how a complex spectrum can arise. As usual in EMFT~\cite{Akerlund:2014mea},
we fix the effective quadratic couplings $\Delta_i$ by matching the bilinear expectation values to an approximation to the point-to-point
correlator of the full model,
\begin{equation}
  \label{eq:dyson_emft}
  G_{\text{EMFT},c} = \int\rd^dk\,G_c(k) = \int\rd^dk\,\left[G_{0,c}^{-1}(k)+\Sigma(k)\right]^{-1} \approx
  \int\rd^dk\,\left[G_{0,c}^{-1}(k)+\Sigma_\text{EMFT}\right]^{-1}.
\end{equation}
This is a matrix equation where $G_{0,c}(k)$ is the connected Green's function of the free theory. It is not immediately clear what the free theory of
a spin model is, but the approximation above is in fact valid for any choice. A good choice will be close to the model we want to study
and at the same time allow for an efficient numerical treatment. We have chosen the free model to have the same action as the original
$\mathbb{Z}_3$ model, eq~\eqref{eq:S_potts}, but with the variables $P$ ranging freely over the complex plane. With this choice the
free connected Green's function is given by $G_{0,c}^{-1} = -2\beta{}\rm{Id}_2\sum_{\nu}\cos{}k_\nu$. The self-energy $\Sigma(k)$ in eq.~\eqref{eq:dyson_emft}
then arises due to the restriction of the field to take values in $\mathbb{Z}_3$. The EMFT self-energy $\Sigma_\text{EMFT}$ is likewise identified
as the difference between the variance of eq.~\eqref{eq:S_EMFT} with $(\Re{}P,\Im{}P)\in\mathbb{R}^2$ and the variance when $P\in\mathbb{Z}_3$
and is given by $G_{\text{EMFT},c}^{-1}+\Delta$, with
\begin{align}
  G_{\text{EMFT},c} &= 2\begin{pmatrix}\expv{\left(\Re{}P\right)^2} - \expv{\Re{}P}^2 & -i\left(\expv{i\Im{}P\Re{}P}-\expv{i\Im{}P}\expv{\Re{}P}\right)\\
    -i\left(\expv{i\Im{}P\Re{}P}-\expv{i\Im{}P}\expv{\Re{}P}\right) & \expv{\left(\Im{}P\right)^2} + \expv{i\Im{}P}^2 \end{pmatrix},\\
  \Delta &= \begin{pmatrix}\Delta_1 & i\Delta_3\\ i\Delta_3 & \Delta_2 \end{pmatrix}.
\end{align}
Hence, the final self-consistency equation becomes
\begin{equation}
  \label{eq:dyson_emft_2}
  G_{\text{EMFT},c} = \int\rd^dk\,\left[G_{\text{EMFT},c}^{-1}+\Delta-2\beta{}\rm{Id}_2\sum_\nu\cos{}k_\nu\right]^{-1}.
\end{equation}
It is clear that $G_{\text{EMFT},c}^{-1}+\Delta-2\beta{}\rm{Id}_2 \equiv \beta M$ plays the role of a mass matrix and we should diagonalize it to obtain
the mass spectrum. It will also be vastly more efficient to integrate over the momenta when $M$ is diagonal. The mass matrix can be parametrized as
\begin{equation}
  \label{eq:Mass_matrix}
  M = \begin{pmatrix} a+b & ic\\ic&a-b\end{pmatrix},
\end{equation}
where $a,b,c\in\mathbb{R}$. The eigenvalues are then given by $m_\pm=a\pm\sqrt{b^2-c^2}$, such that if $\abs{c}>\abs{b}$ the spectrum will consist
of a pair of complex conjugated masses $m_R\pm im_I$ with $m_R = a$ and $m_I = \sqrt{c^2-b^2}$. This implies cosine-modulated exponential fall-off
in the correlators in the $\left(\Re{}P,\Im{}P\right)$ basis, as expected. By solving the model in the $(h_R,h_I)$-plane, a phase diagram
analogous to what was obtained in one dimension with the transfer matrix, Fig.~\ref{fig:pd_1d_z3_tm}, can be constructed by studying the
behavior of the masses. In Fig.~\ref{fig:pd_3d_z3_emft} we show the results for fixed $\beta=0.08$, with the most interesting features being the
disorder lines (in red), where the masses are degenerate, and the blue dashed lines where the real part of the complex masses vanishes. Beyond these lines
the momentum
integral in the self-consistency equation no longer converges, since the integrand is no longer decaying at large distances. 
One may guess that with purely imaginary masses, the system would enter a crystalline phase with a purely trigonometric correlator but
there is no way to verify that using EMFT. This phase diagram can then be compared both to the mapping
$(h_R,h_I)\to(e^{-M/T}\cosh(\mu/T),e^{-M/T}\sinh(\mu/T))$ and to the alternative mapping in~\cite{Condella:1999bk}. It is found that the second case
covers a subspace of $M,\mu\in\mathbb{R}$ and there are indeed regions in parameter space where the mapping is valid and where one expects
a complex spectrum. However, in that region $h_R$ is substantially larger than $h_I$ which means in the $M,\mu$ variables that both $M$ and $\mu$
are rather small, which is presumably far away from the region where the model is expected to be a valid approximation of QCD.

Now that an approximate phase diagram has been obtained, we can select points in the liquid phase which are favorable in terms of $m_R$
and $m_I$ where full Monte Carlo simulations using the worm algorithm will be performed.

\section{Results}\noindent
\label{sec:results}
For our lattice simulations we used the flux-variables representation described in~\cite{Mercado:2012yf}, with quark occupation number
$n_x\in\{-1,0,1\}$ on each site and flux occupation number $l_{x,\nu}\in\{-1,0,1\}$ on each link. Gauss' law requires that the flux at each
site is a multiple of three. Allowed configurations consist of flux-tube networks with or without attached quarks. If there are no quarks
attached the flux-network can be thought of as a glueball. There are also neutral networks with any number of quarks and an equal number
of anti-quarks attached, for example networks connecting one quark with an anti-quark can be thought of as mesons. The third possibility
is to have a surplus of $3n$ (anti-)quarks. This is equivalent to having the junctions of the network to sum up to $n$, we say that the
network has junction charge $n$. These charged networks are associated with baryons.

\begin{figure}[htp]
\centering
\includegraphics[width=0.49\linewidth]{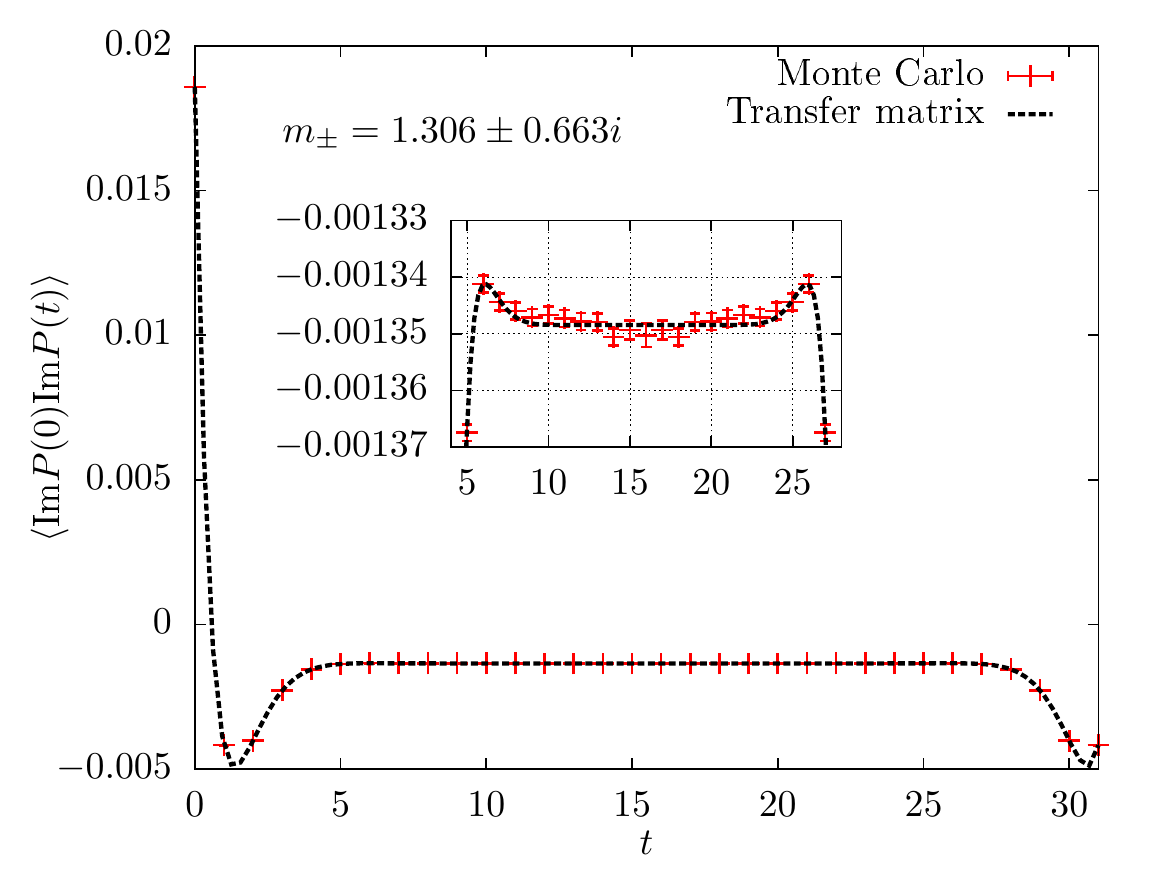}
\includegraphics[width=0.49\linewidth]{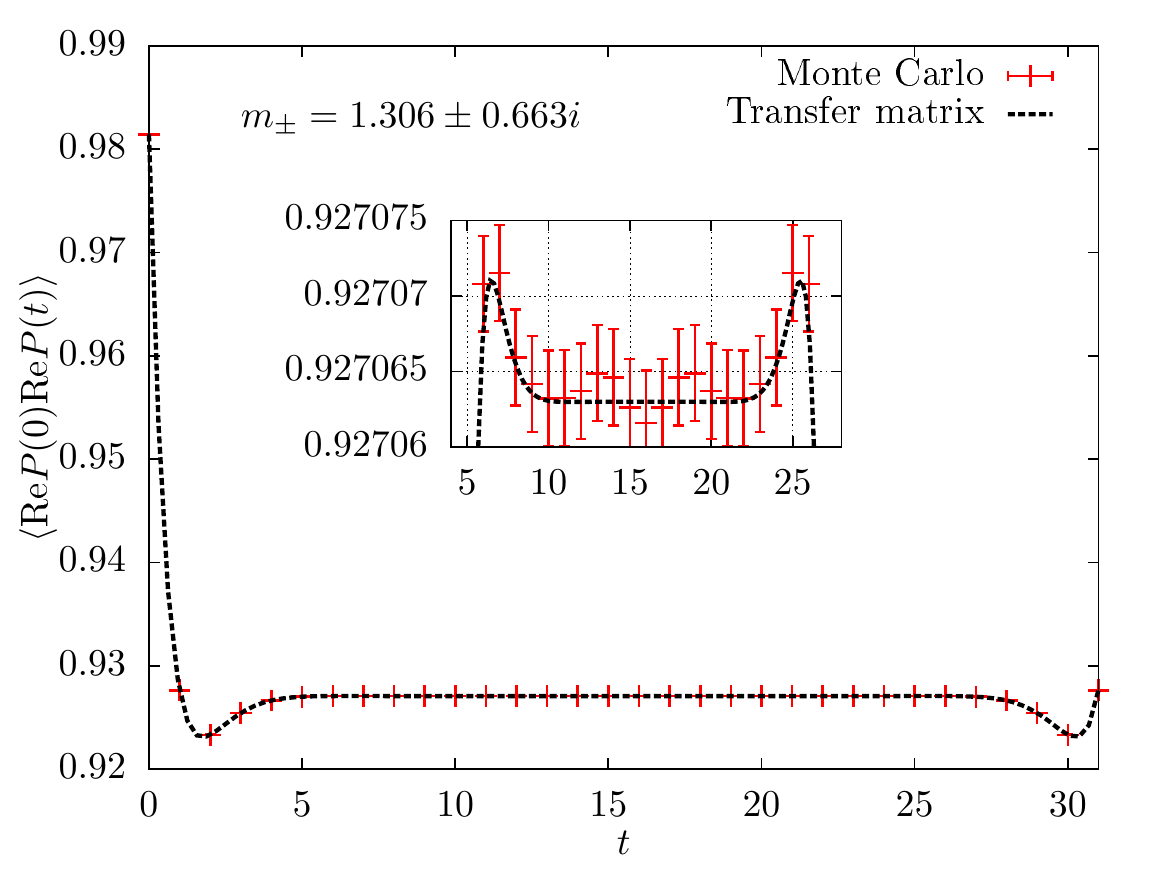}
\caption{Two components of the spin-spin correlator in one dimension for $\beta=0.5,e^{-M/T}=0.02$ and $\mu/T=3.689$. There is a clear oscillation
  in both correlators and the result agrees perfectly with the exact result obtained using the transfer matrix. The complex mass is
  given by $m_\pm \approx 1.306\pm0.663i$.}
  \label{fig:1d_corr}
\end{figure}

The worm algorithm generates a Markov chain of allowed configurations by temporarily violating the constraint, something which can be
exploited to obtain improved estimators for spin-spin correlation functions. In addition to the usual $\expv{P(0)P^\dagger(x)}$ we use
a modification introduced in~\cite{Rindlisbacher:2015xku,Rindlisbacher:2016zht}, which allows us access to improved estimators of also
$\expv{\Re{}P(0)\Re{}P(x)}$ and
$\expv{\Im{}P(0)\Im{}P(x)}$. This is crucial because the best signal-to-noise ratio will be found in the correlator of the imaginary
parts of the spins, since it has the smallest constant background. We first reproduced the results obtained by the transfer matrix
method in one dimension in order to verify that the algorithm was properly implemented. A typical correlator in the liquid phase is
shown in Fig.~\ref{fig:1d_corr} and there is perfect agreement with the analytic result for all three propagators. It should be noted that
the real part of the
mass $m_R$ is in general always large when the imaginary part $m_I$ is of order one or larger, this makes it very difficult to resolve
the first local maximum of the correlator.

\begin{figure}[htp]
\centering
\includegraphics[width=0.6\linewidth]{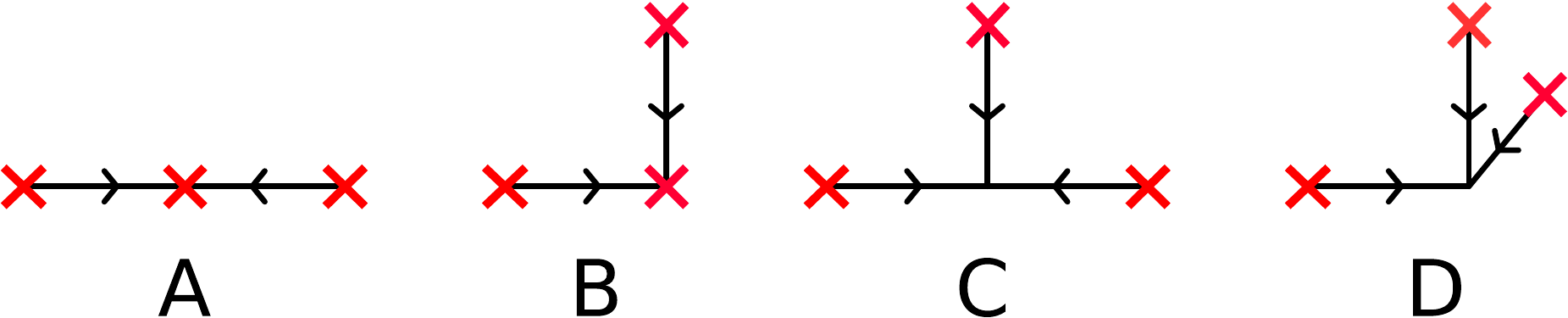}
\caption{The different junctions allowed in the flux-variable representation of the $\mathbb{Z}_3$ model described in the text. The red crosses
  represent quarks and the lines represent the directed flux-tubes. The junction is located in the center of each network where the flux sums up to
  three. Note that the quarks bounding the network may also be replaced by arbitrary larger networks of charge one. In one dimension only junction
  A is possible. The three-dimensional junction D is only present in dimension three or higher.}
  \label{fig:junctions}
\end{figure}

\begin{figure}[htp]
\centering
\includegraphics[width=0.49\linewidth]{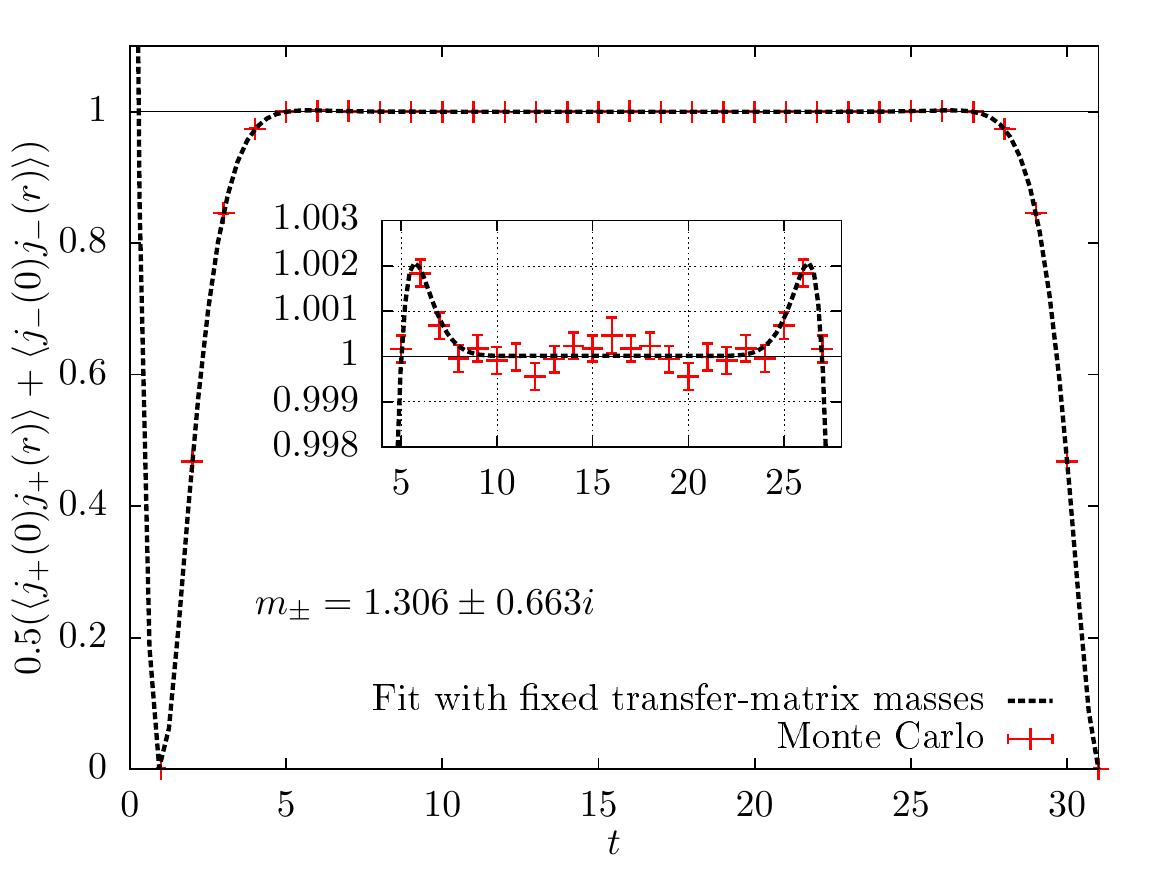}
\includegraphics[width=0.49\linewidth]{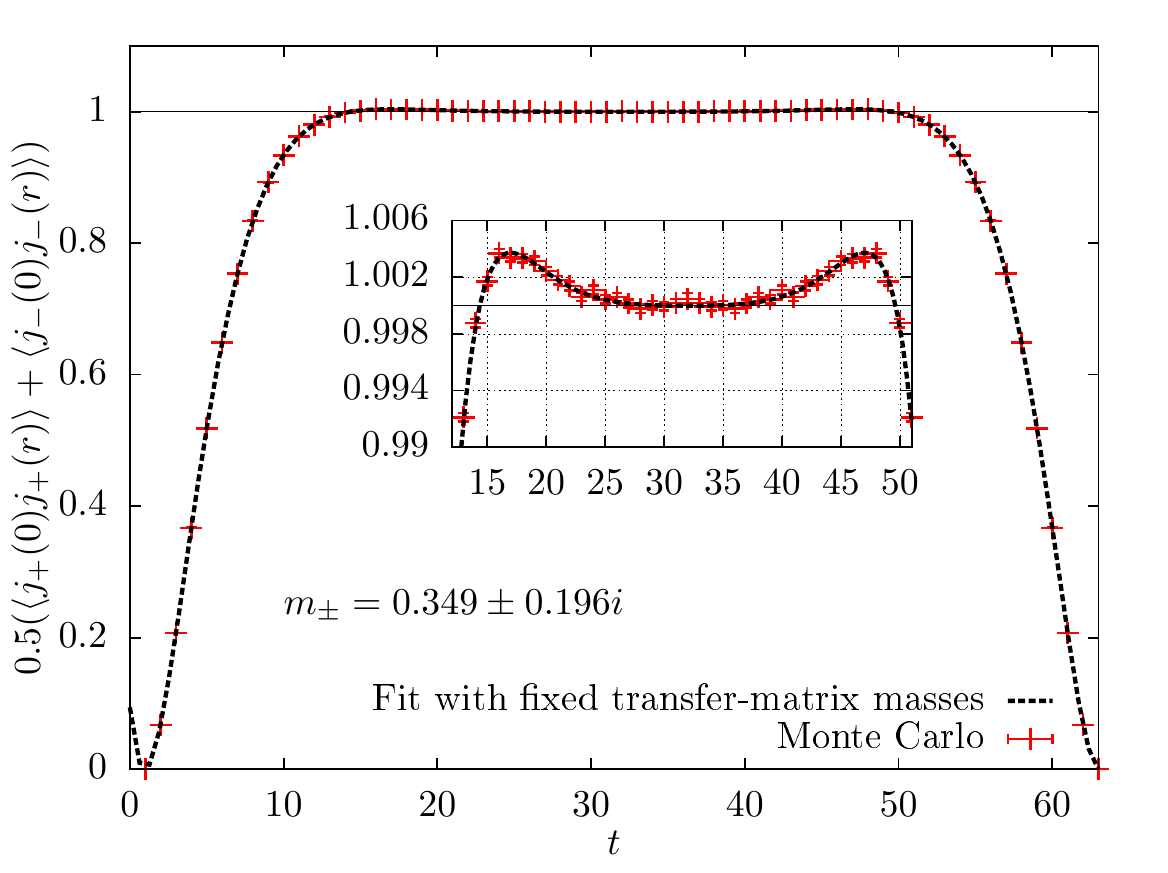}
\caption{The junction-junction correlator in one dimension for $\beta=0.5,e^{-M/T}=0.02$ and $\mu/T=3.689$ (\emph{left panel}) and for
  $\beta=1.2,e^{-M/T}=0.0042$ and $\mu/T=4$ (\emph{right panel}). The signal of oscillation is even clearer than in the spin-spin correlator
  since this observable is less noisy, cf. Fig.~\ref{fig:1d_corr}. The fits are given by eq.~\eqref{eq:corr_exact} with the mass fixed at
  the value obtained with the transfer matrix.}
  \label{fig:1d_junction}
\end{figure}

We also measured the junction-junction correlator on the configurations generated by the worm algorithm. The junction $j_x$ takes the value
$n$ if $3n,\,n\in\mathbb{Z}$ units of flux flow into the site $x$. With the flux
variables described above there are in general four types of junctions, depicted in Fig.~\ref{fig:junctions} but in one dimension only junction A
with one quark and two in-going fluxes (or its reverse) attached to the site is possible. In Fig.~\ref{fig:1d_junction} we show the correlation
between positive $j_+$ and negative $j_-$ junctions for two different parameter values. Here the oscillation is even clearer due to a less noisy
observable, although we do not have an improved estimator for this correlator. The dashed line is obtained by fitting the amplitude and phase in
eq.~\eqref{eq:corr_exact} while keeping the masses fixed at the exact values obtained by the transfer matrix. The mass is the same as for the
spin-spin correlator since the junction is a local object and there is only one (complex) mass in the one-dimensional case. It should be noted that these
parameter values have been selected to give a maximally clear first maximum in the oscillation. For general parameter values in the liquid phase
it is only possible to see the first minimum, while the first maximum is drowned in noise. This will be especially true in three dimensions where
the real part of the mass is larger than in the one-dimensional case.

\begin{figure}[htp]
\centering
\includegraphics[width=0.49\linewidth]{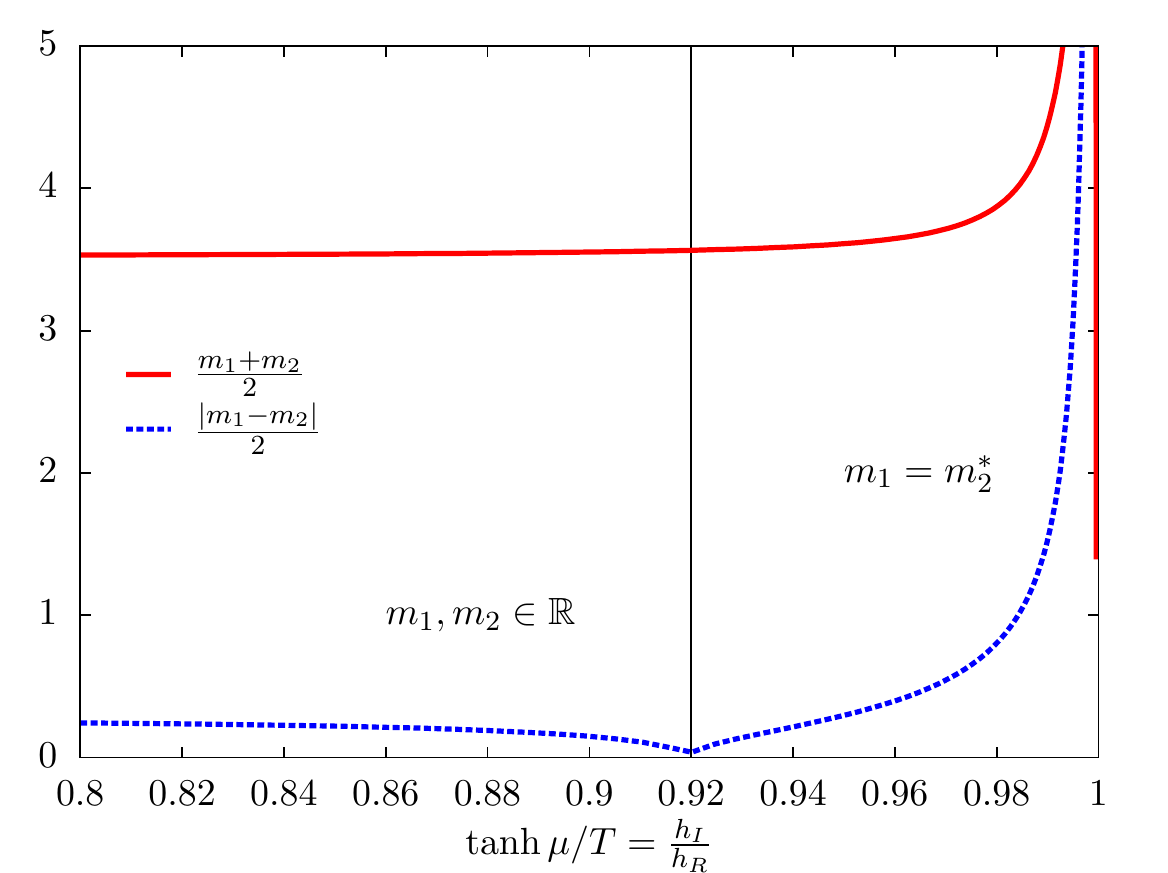}
\includegraphics[width=0.49\linewidth]{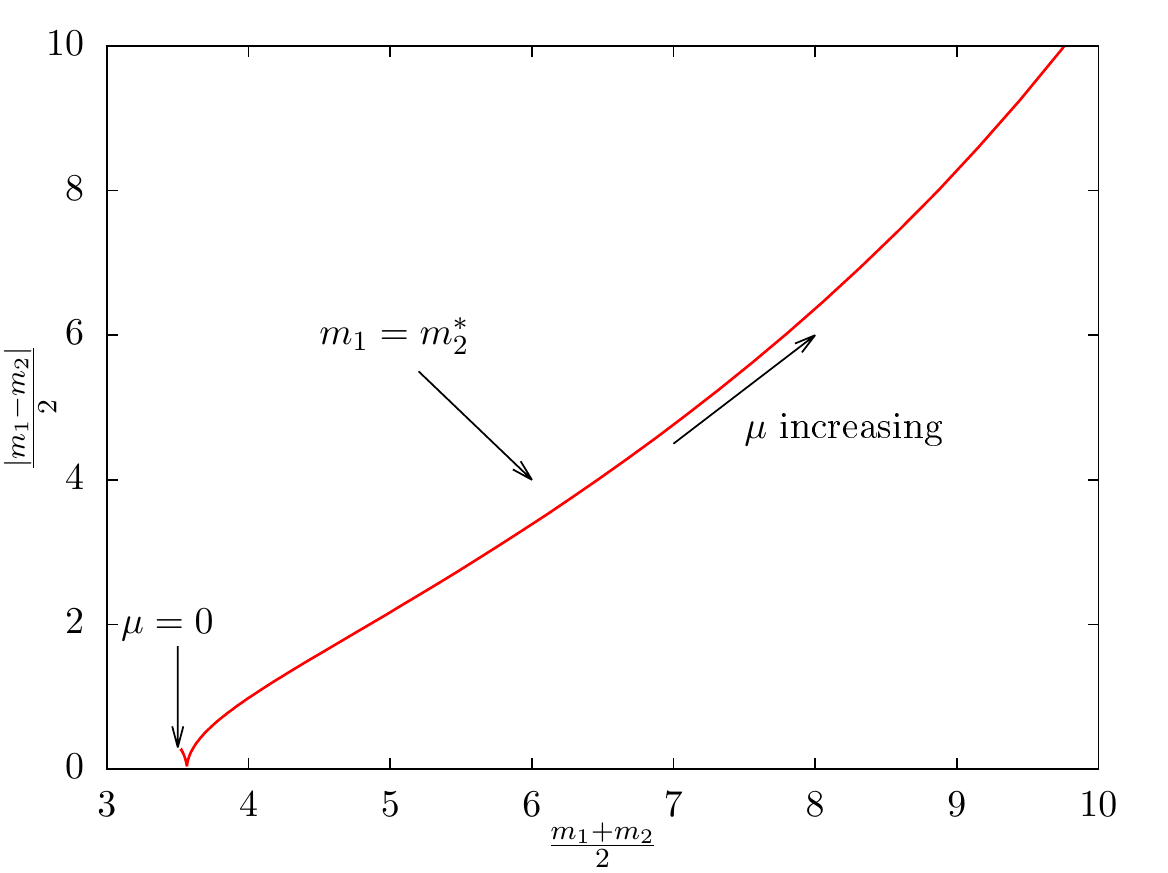}
\caption{\emph{Left}: the masses obtained by EMFT as a function of $\tanh\mu/T=h_I/h_R$ for fixed $\beta=0.08$ and $e^{-M/T}=0.05$.
  \emph{Right}: imaginary part versus real part of the complex mass, for the same
  parameters as in the left panel. The feature that the real part rises approximatively linearly with the imaginary part is generic,
  as is the fairly large value of the real part at $\mu=0$. In the part of the curve to the left of the cusp, both masses are real and
  their half-difference is shown as a function of their half-sum.}
  \label{fig:masses_emft}
\end{figure}

\begin{figure}[htp]
\centering
\includegraphics[width=0.6\linewidth]{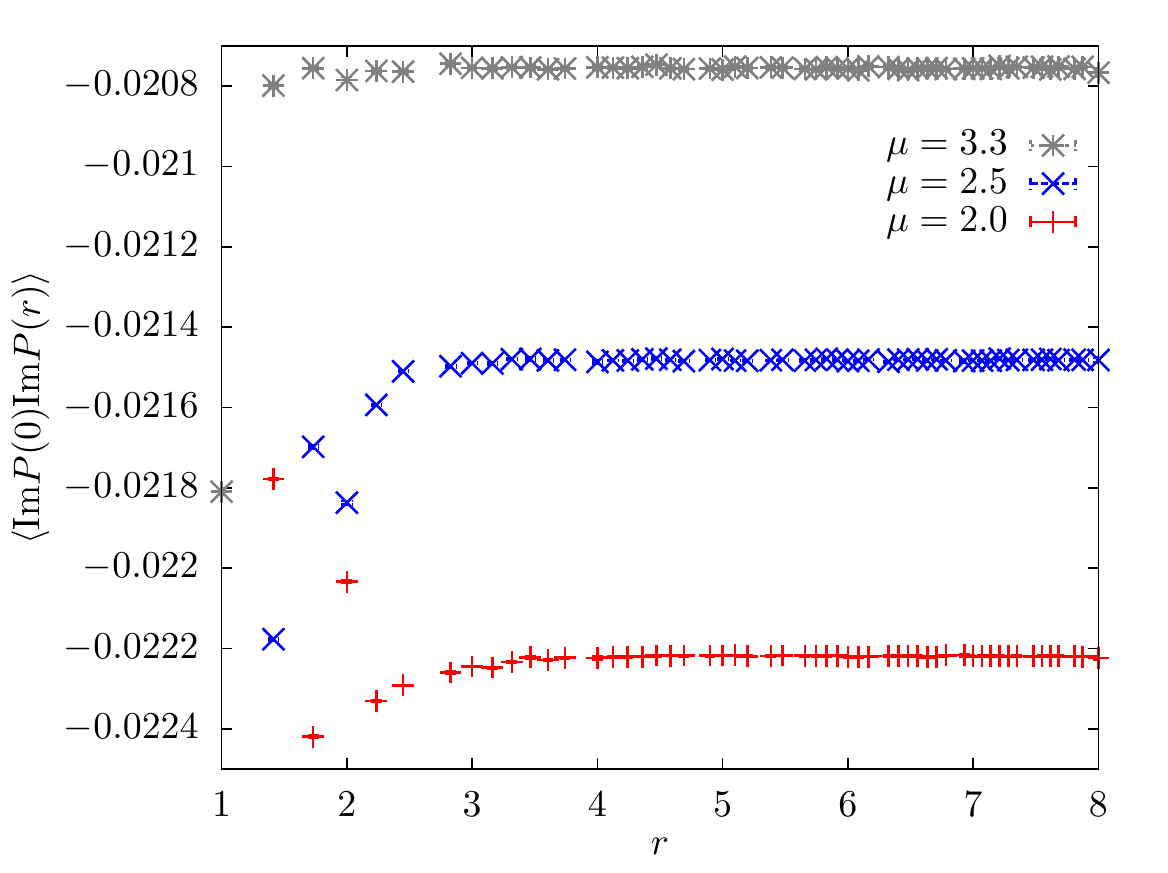}
\caption{The correlator of the imaginary part of the spins for three different chemical potentials $\mu$ at fixed $\beta=0.08$ and $e^{-M/T}=0.05$.
  The minimum of the correlator moves towards zero and its width decreases as $\mu$ increases, suggesting that the imaginary part of the
  mass increases with $\mu$, as expected. The significant staggered contribution to the correlator makes a fit to the data difficult.
  The data sets are shifted vertically for clarity and the data points at $r\leq1$ are far above the shown data points, i.e. we have zoomed
  in on the minimum of the correlators.}
  \label{fig:3d_corr}
\end{figure}

We then move to the physically interesting case of three
dimensions, and guided by the phase diagram calculated by EMFT we select a few points assumed to be in the liquid phase and look
for the corresponding signals in the correlators. Also here, however, the damping of the correlator is always strong, as is illustrated
in Fig.~\ref{fig:masses_emft}. In the left panel $m_R$ and $m_I$, obtained by EMFT, are plotted as a function of $\tanh\mu/T$ for fixed $M/T$ and $\beta$
and in the right panel $m_I$ is plotted as a function of $m_R$ to emphasize the approximately linear growth relation. Both masses increase with $\mu/T$.
As a consequence, it is typically only possible to resolve the first minimum of the oscillating correlator. In Fig.~\ref{fig:3d_corr}
we show correlators of $\Im{}P$, as a function of the Euclidean distance $r=\sqrt{x^2+y^2+z^2}$, obtained by our worm simulations for three different
values of the chemical potential $\mu/T\in\{2.0,2.5,3.3\}$
at fixed $\beta=0.08$ and $e^{-M/T}=0.05$. There is a clear staggered component in the correlators, which makes it very hard to fit the data
to a simple ansatz. This short-distance effect, whose sign is $\mu$-dependent (Fig.~\ref{fig:3d_corr}, blue vs red), stems from the size and
shape of the junctions, shown in Fig.~\ref{fig:junctions}: in $d=1$,
only A is possible. For all $\mu$ there is a clear minimum whose position moves toward zero and whose width decreases
as the chemical potential increases. This suggests that the imaginary part of the mass increases with $\mu$, as it does in one dimension and
as EMFT predicts. 
Also, in neither of these correlators is it possible to see a maximum. This is not very surprising, but a discernible maximum would
be indisputable evidence of a complex spectrum.

\begin{figure}[htp]
\centering
\includegraphics[width=0.49\linewidth]{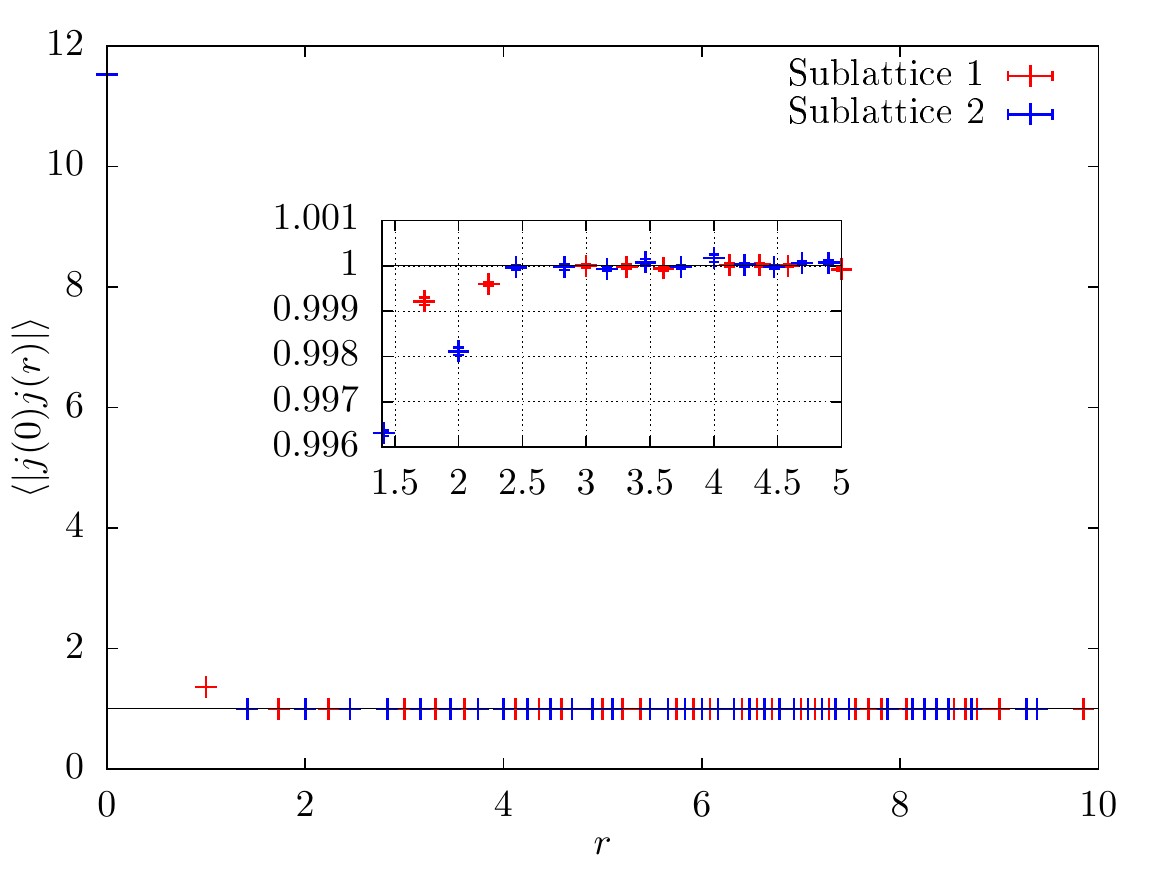}
\includegraphics[width=0.49\linewidth]{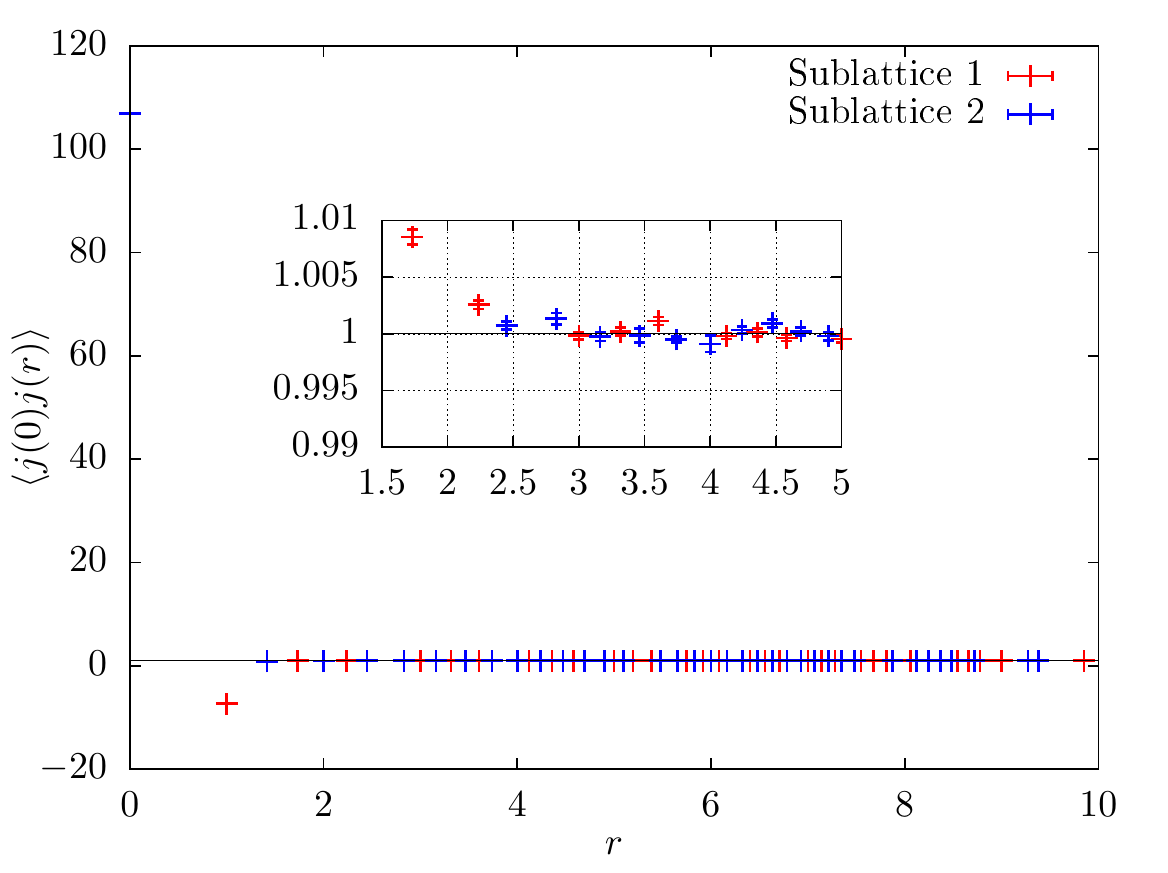}
\caption{The correlator of the absolute value of the junction number (\emph{left panel}) and the correlator of the junction number
  (\emph{right panel}) for $\beta=0.08,e^{-M/T}=0.05$ and $\mu/T=3.3$ on a $12^3$ lattice. The depletion in the left correlator and the
  enhancement in the right correlator around distance $2$ support the proposition that the system behaves as a liquid. However,
  the strong staggered character still leaves some doubt. In the right panel, the data point at $\sqrt{2}$ and $2$ are far below $0.99$
  and are omitted such that it can be clearly seen that the points at $\sqrt{3}$ and $\sqrt{5}$ are above $1$.}
  \label{fig:3d_jct}
\end{figure}

If the mass spectrum is complex and the system behaves like a liquid then the junction-junction correlator should also show the characteristic,
oscillatory behavior seen in the $1d$ model. Since the junction-junction correlator is less noisy than the spin-spin one, one may even hope
that a maximum of the oscillating correlator can be resolved, thus establishing the complex spectrum without doubt.
In Fig.~\ref{fig:3d_jct} we show two junction-junction correlators for $\beta=0.08,e^{-M/T}$ and $\mu/T=3.3$. In the left panel we show the
correlator of the absolute values of the junctions whereas in the right panel the sign of the junctions is also taken into account.
The difference of scales of the two correlators comes from the fact that they are both normalized to one at large distances and that
$\expv{\abs{j}}\sim3\expv{j}$. To emphasize the staggered component of the correlators we plot the correlators on the two different
sub-lattices with different colors. Inspecting first the charge-insensitive correlator (\emph{left panel}) we see that there is indeed
a depletion in the density of junctions of any type within distance $[1,2.5]$ of a junction but it is not possible to tell if this minimum
is followed by a maximum. In the charge-sensitive correlator (\emph{right panel}) there is a clear maximum in the correlator in roughly the
same interval, but only in one of the sub-lattices. This strong staggered dependence is of course a lattice artifact. It should however be
noted that for smaller values of $\mu/T$ (and thus longer wavelength oscillations, cf. Fig~\ref{fig:3d_corr}) there is a clear, broader, minimum
in both correlators and on both sub-lattices, which indicates that the effect is not merely a staggered effect, although the maximum which is
predicted to follow is completely damped away. All in all, the behavior of the different correlators strongly suggests that there is a complex
mass spectrum at the investigated parameter values, and that the prediction in~\cite{Nishimura:2015lit} that the phase structure observed in
one-dimension has an analogue also in three dimensions is correct.

\begin{figure}[htp]
\centering
\includegraphics[width=0.6\linewidth]{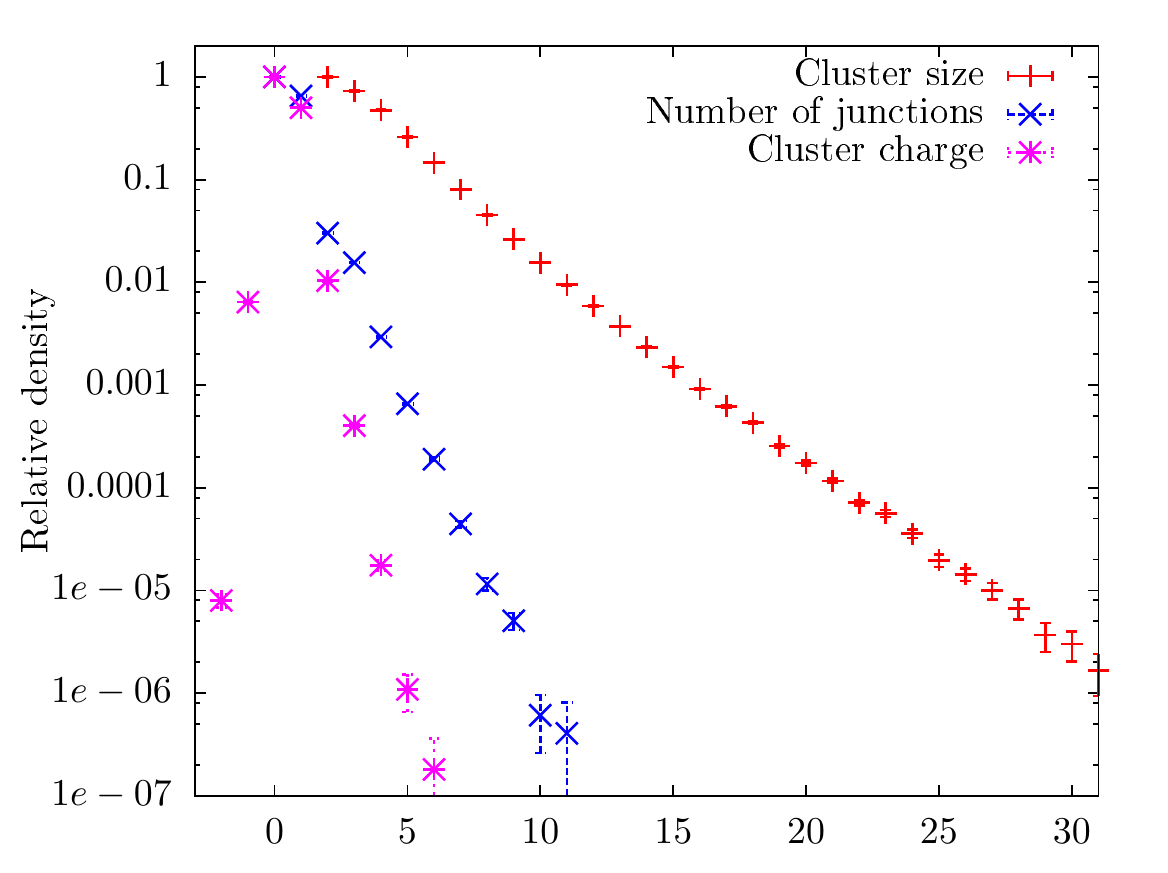}
\caption{Histograms of the distribution of flux-network size, the number of junctions in a network
  and the network charge for $\beta=0.08,e^{-M/T}=0.05$ and $\mu=2.0$ on a $12^3$ lattice.}
  \label{fig:cluster_stats}
\end{figure}

Finally, we measured some statistics of the flux-tube networks and the junctions. Using the labeling of Fig.~\ref{fig:junctions} we
find that the ratio of C to D junctions is very close to $3/2$ and the ratio of B to A junctions is very close to 4, both in full
agreement with entropic arguments, i.e. their relative abundance is obtained by counting the number of possible different orientations
for each type of junction, assuming all orientations appear with equal probability.
The ratio of pure-flux junctions (C\&D) to flux-quark junction (A\&B) depends on the parameters but
for the parameter values we used the flux-quark junctions typically outnumber the pure-flux junctions by a factor 10, reflecting the energy cost
of the additional flux tube.
In Fig.~\ref{fig:cluster_stats} we show the histograms of the distribution of the flux-network size, the number of junctions in a network
and the network charge for $\beta=0.08,e^{-M/T}=0.05$ and $\mu=2.0$ on a $12^3$ lattice.

\section{Conclusions}\label{sec:conclusions}\noindent
Using unbiased Monte Carlo simulations, we have shown that the $\mathbb{Z}_3$ spin model in three dimensions with charge-symmetry breaking
external fields has non-monotonic correlators, both in the original spin variables and in the flux variables, for some regions in parameter
space. This strongly suggests that the spectrum in these regions is complex and this claim is also backed up by EMFT calculations of the
three-dimensional model. Of special interest is the oscillatory nature of the
junction-junction correlator, which is the analogue of the baryon-baryon correlator in heavy-dense QCD, for some regions of parameter space.
The possibility that models with complex saddle points may have a complex mass spectrum and thus non-monotonic correlators has previously been established
analytically in the one-dimensional case
and has been argued to also hold true in higher dimensions. We have shown that the worm algorithm is capable of reproducing these results, even though
the original spin model suffers from a strong sign problem. The phase diagram in one dimension contains regions where the system behaves like
a liquid, with exponentially damped oscillations, and like a crystal with a purely oscillatory correlator. In general, it is
expected~\cite{Nishimura:2015lit} that these features carry over also to higher dimensions. For those regions of
parameter space where the model has a sign-problem free representation we have only found evidence of the liquid phase with exponentially
damped but oscillating correlations between spins, as well as junctions. We have found no evidence of a crystalline phase and it is probable that it lies
beyond the reach of the worm algorithm in three dimensions, as it does in one dimension. It should also be noted that even the liquid phase
may lie in an unphysical region of parameter space. At least according to EMFT it lies outside of the region of parameter space which
can be mapped to the more physical flux-tube model of Condella and Detar~\cite{Condella:1999bk}. A complex mass spectrum can only be found in a
parameter region where the mass $M$ of the underlying heavy quark satisfies $M\ll T$, whereas the validity of the effective description of QCD
by a Potts model requires $M\gg T$. This situation may change if the junctions are given a nonzero weight as in~\cite{Patel:2011dp}, but this
possibility has not been investigated here. However, whatever the values of the other parameters, introducing a junction weight will further damp the
signal we want to measure, making the search yet more difficult.

Our findings supports the claim that in general, it is plausible that models without charge-conjugation symmetry, but invariance under the combined
action of charge conjugation and complex conjugation, will have regions with a complex mass spectrum in their phase diagram. However, more work is
needed before precise statements can be made about whether or not this is a phenomenon which occurs under physical conditions.
This first proof of principle should encourage the study of more realistic models, and the search for experimental signals in heavy-ion
collisions as advocated in~\cite{Patel:2011dp}.

\bibliographystyle{JHEP.bst}
\bibliography{paper_oscillating.bib}

\end{document}